\documentclass[tradiabstract]{aa} 
\usepackage{graphicx}
\usepackage{txfonts}
\newcommand{\mic}{$\mu$m}
\newcommand{\ang}{$\rm \AA$}
\newcommand{\halpha}{H$\alpha$}

\newcommand{\Msun}{M$_{\odot}$}

\newcommand{\teff}{$T_{\mathrm{eff}}$}

\begin{document}

   \title{MBM 12: young protoplanetary discs at high galactic latitude}


   \author{G. Meeus\inst{1}
          \and
          A. Juh\'asz\inst{2}
          \and
          Th. Henning\inst{2}
          \and
          J. Bouwman\inst{2}
          \and
          C. Chen\inst{3}
          \and
          W. Lawson\inst{4}
          \and
          D. Apai\inst{3}
           \and 
          I. Pascucci\inst{5} 
          \and 
          A. Sicilia-Aguilar\inst{2}
          }

   \institute{Astrophysical Institute Potsdam, 
              An der Sternwarte 16, D-14482 Potsdam, Germany\\
              \email{gwen@aip.de}
         \and
             Max Planck Institute for Astronomy, K\"onigstuhl 17, D-69117 
             Heidelberg, Germany\\
         \and
             Space Telescope Science Institute, 3700 San Martin Dr.,
             Baltimore, MD 21218, USA\\
         \and
             School of Physical, Environmental and Mathematical Sciences,
             University of New South Wales, Australian Defence Force Academy,
             ACT 2600, Canberra, Australia
         \and
             Department of Physics and Astronomy, John Hopkins University,
             Baltimore, MD 21218, USA\\
             }

   \date{Received 2008; accepted  }

 
  \abstract
   {We present {\it Spitzer} infrared observations to constrain disc and dust
   evolution in young T Tauri stars in MBM 12, a star-forming cloud at high
   latitude with an age of 2 Myr and a distance of 275~pc. The region contains
   12 T Tauri systems, with primary spectral types between K3 and M6; 5 are
   weak-line and the rest classical T Tauri stars. We first use MIPS and
   literature photometry to compile spectral energy distributions for each of
   the 12 members in MBM 12, and derive their IR excesses. Of the 8 stars that
   are detected with MIPS (spectral types between K3 and M5), only 1 lacks an
   IR excess - the other 7 all have an IR excess that can be attributed to a
   disc. This means that in MBM 12, for the detected spectral types K3-M5, we
   have a very high disc fraction rate, about 90\%. Furthermore, 3 of those 7
   excess sources are candidate transitional discs. The four lowest-mass
   systems in the cloud, with spectral types of M5-M6, were undetected by {\it
   Spitzer}. Their upper limits indicate that they either have a
   transitional disc, or no disc at all. The IRS spectra are analysed with the
   newly developed two-layer temperature distribution (TLTD) spectral
   decomposition method. For the 7 T Tauri stars with a detected IR excess, we
   analyse their solid-state features to derive dust properties such as
   mass-averaged grain size, composition and crystallinity. The mass-averaged
   grain size we determine from the 10 micron feature has a wide range,
   between 0.4 and 6~\mic. This grain size is much smaller in the
   longer-wavelength region: between 0.1 and 1.5~\mic.  We find that
   later-type objects have larger grain sizes, as was already shown by earlier
   studies. Furthermore, we find a wide range in mass fraction of the
   crystalline grains, between 3 and (at least) 30\%, with no relation to the
   spectral type nor grain size.  We do find a spatial gradient in the
   forsterite to enstatite range, with more enstatite present in the warmer
   regions.  The fact that we see a radial dependence of the dust properties
   indicates that radial mixing is not very efficient in the discs of these
   young T Tauri stars. The sources that have the least amount of disc flaring
   have the largest grain sizes, pointing to dust settling. A comparison
   between the objects with companions closer than 400 AU ('binaries') and
   those with wider or no companions ('singles'), shows that disc evolution
   already starts to differentiate between both groups at an age of 2 Myr: the
   excess at 30~\mic \ is a factor 3 larger for the 'single' group.
   The SED analysis shows that the discs in MBM 12, in general, undergo rapid
   inner disc clearing, while the binary sources have faster disc
   evolution. The dust grains seem to evolve independently from the stellar
   properties, but are mildly related to disc properties such as flaring and
   accretion rates.
   }


   \maketitle

%

\section{Introduction}

Our understanding of the star and planet formation process has advanced
through comprehensive observations of nearby star-forming regions, such as the
Taurus-Auriga cloud or the much denser and massive Orion Nebula Cluster.
Because these sites are relatively close and young, it is often possible to
determine the complete census of cluster members, from the highest mass stars
down to brown dwarfs with masses of just a few Jupiters (e.g. Luhman et al.\
\cite{luhman2003}, Muench et al.\ \cite{muench2002}). Furthermore, the
detection of companions through adaptive optics and other high-resolution
imaging and the subsequent derivation of multiplicity rates is an important
piece of the star formation (SF) puzzle (e.g. Petr et al.\ \cite{petr1998},
Ratzka et al.\ \cite{ratzka2005}). In addition, near-IR studies of SF regions
have shown that young objects exhibit near-IR excesses, commonly attributed to
the presence of protoplanetary discs (e.g. Adams et al.\ \cite{adams1987},
Lada et al.\ \cite{lada2000}). The structure, mass and size of those discs
provide constraints on the process of planet formation in these young
environments.

To study the young cluster members in more detail, it is important to 
characterise their circumstellar environment, which is best done at infrared
and/or millimetre wavelengths. With the launch of {\it Spitzer}, this field
has made huge steps forward, as for the first time, it was possible to observe
the disc and dust characteristics of large samples of T Tauri stars
(e.g. Kessler-Silacci et al.\ \cite{kessler2006}, Sicilia-Aguilar et al.\
\cite{aurora2007}, Pascucci et al.\ \cite{pascucci2008}) and even brown dwarfs 
(Apai et al.\ \cite{apai2005}, Scholz et al.\ \cite{scholz2007}).
 
Cloud 12 in the catalogue of Magnani, Blitz and Mundy (\cite{magnani1985}), in
short MBM 12, is a molecular cloud at high galactic latitude ($l, b$ =
159.4$\degr$, -34.3$\degr$), with relatively high extinction (A$_{\mathrm{V}}
> 5$ mag). Radio maps of the region in CO give a mass estimate for the whole
complex of 30 to 200 \Msun \ (Pound et al. \cite{pound1990}). The cloud does
not appear to be gravitationally bound, and the molecular gas is expected to
dissipate within the next few Myrs (Zimmermann \& Ungerechts
\cite{zimmermann1990}).  Hearty et al.\ (\cite{hearty2000}) used {\it ROSAT}
observations to detect X-ray sources in MBM 12, and followed-up stellar
candidates by optical spectroscopy, to confirm 8 X-ray emitting T Tauri stars
with an upper age of 10~Myr. It is one of the few clouds at high galactic
latitude known to harbour such young objects.  From its content, the cloud can
be seen as a precursor of a TW Hydrae-like association, where the molecular
material has not yet disappeared.  Although MBM 12 was initially thought to be
one of the nearest star formation regions, at a distance of only 65~pc, it is
now determined to lie at a distance of 275~pc (Luhman \cite{luhman2001}).

A census of this association - complete down to masses of 0.03~\Msun \ - was
determined by Luhman (\cite{luhman2001}), based on sensitive near-IR and
optical photometry. The candidate members were confirmed by follow-up
spectroscopy, which also allowed to study their H$\alpha$ and Li 6707~\ang \
line properties. In total, 12 T Tauri stars were found to be real members of
MBM 12, and an age of 2$^{+3}_{-1}$ Myr was derived from both the lithium line
and the location of the objects in the H-R diagram. The spectral types for
those 12 TTS is between K3 and M6. An additional study by Luhman \& Steeghs
(\cite{luhman2004}) of 7 candidate members could not confirm any more members.
The T Tauri stars in MBM 12 have a high binary frequency: near-infrared
adaptive optics studies by Chauvin et al.\ (\cite{chauvin2002}) and Brandeker
et al.\ (\cite{brandeker2003}) revealed that MBM 12 contains at least 4
binaries and 2 triples with projected separations between 20 and 4000 AU (of
which one is a candidate quadruple: Lk\halpha~262 has a projected distance of
only 15~arcsec to the triple Lk\halpha~263).

In this paper, we present {\it Spitzer} IRS and MIPS observations of all MBM
12 members, and combine our observations with literature photometry and
spectroscopy to analyse the derived dust characteristics by relating them to
the stellar parameters and disc properties.  In Sect.~\ref{sample}, we
present the individual targets and the {\it Spitzer} observations. The
analysis in Sect.~\ref{analysis} first discusses the spectral energy
distributions, and then the spectral features are interpreted in terms of dust
properties with the aid of a spectral decomposition model. In
Sect.~\ref{discussion}, we discuss grain growth, crystallisation, disc
properties and accretion rates. We round off with conclusions in
Sect.~\ref{conc}, and give more details in the Appendix on the dust model 
used and the mass fractions derived.

\section{Sample and Observations}
\label{sample}

\subsection{Targets}
\label{s_targets}

\begin{table*}
\begin{minipage}[t]{\textwidth}
\caption{Target coordinates and parameters: Spectral Type, \teff, 
  H$\alpha$ equivalent width of the primary from Luhman (\cite{luhman2001}), T
  Tauri Class derived by us (see Sect.~\ref{s_targets}), multiplicity (S:
  single, B: binary, T: triple and Q: quadruple) and projected separations
  from Chauvin et al.\ (\cite{chauvin2002}) and Brandeker et al.\ 
  (\cite{brandeker2003}). A negative equivalent width means emission; the 
  width is given in angstroms, \AA.}
\label{tsample}     
\centering                        
\renewcommand{\footnoterule}{}  
\begin{tabular}{llcccccccc}       
\hline\hline\noalign{\smallskip}              
Object&Alternative Name
&$\alpha$(2000.0)&$\delta$(2000.0)&Spectral&\teff &Eq. Width&Class&Multiplicity&Projected\\
   &                 &(h m s)         &(\degr~\arcmin~\arcsec)&Type&(K)&
   (\halpha)& &Status&Separation (\arcsec)\\
\noalign{\smallskip}
\hline\noalign{\smallskip}
MBM 12-1  &RX J0255.4+2005  & 02 55 25.78 &20 04 51.7      &K6     &4205&-1  &
WTTS &B  &0.533\\
MBM 12-2  &Lk\halpha \ 262  & 02 56 07.99 &20 03 24.3      &M0     &3850&-40 &
CTTS     &(Q)\footnote{Possible companion to Lk\halpha \ 263 ABC}&15.3\\
MBM 12-3  &Lk\halpha \ 263 (A-B/B-C)& 02 56 08.42 &20 03 38.6&M3 &3415  &-25 &
 CTTS  &T  &0.416/4.1\\
MBM 12-4  &Lk\halpha \ 264  & 02 56 37.56 &20 05 37.1      &K5     &4350&-18 &
CTTS &B  &9.160\\
MBM 12-5  &E 02553+2018        & 02 58 11.23 &20 30 03.5      &K3   &4660 &-3 &  
W/CTTS\footnote{On the border between weak-line and classical TTS, see target notes in Sect.~\ref{s_targets}} &B  &1.144\\
MBM 12-6  &RX J0258.3+1947     & 02 58 16.09 &19 47 19.6      &M5  &3200  &-29 & 
 CTTS &S  &--\\
MBM 12-7  &RX J0256.3+2005     & 02 56 17.98 &20 06 09.9      &M5.75 &3024&-14 &
WTTS &--  &--\\
MBM 12-8  &--                  & 02 57 49.02 &20 36 07.8      &M5.5 &3058&-120 &
CTTS &-- &--\\
MBM 12-9  &--                  & 02 58 13.37 &20 08 25.0      &M5.75 &3024&-10 &
WTTS &--  &--\\
MBM 12-10 &--                  & 02 58 21.10 &20 32 52.7      &M3.25 &3379&-12 &
WTTS &B  &0.390\\
MBM 12-11 &--                  & 02 58 43.80 &19 40 38.3      &M5.5 &3058&-14 &
WTTS &--  &--\\
MBM 12-12 & S 18 (A-B/Ba-Bb)   & 03 02 21.05 &17 10 34.2      &M3   &3415&-69 &
CTTS & T  &0.747/0.063\\
\noalign{\smallskip}
\hline
\end{tabular}
\end{minipage}
\end{table*}

Our sample is unbiased, as it includes all the confirmed members of the MBM 12
cloud (Luhman \cite{luhman2001,luhman2004}). Their parameters, spectral type
and T Tauri Class (weak-line or classical) are listed in Table~\ref{tsample}.
We re-derived their T Tauri Class, following White and Basri
(\cite{white2003}): T Tauri stars are classical when, for spectral types
between K0 and K5, the equivalent width of their \halpha \ line,
$\mid$EW(\halpha)$\mid \ \geq$ 3 \AA, for K7 to M2.5, $\mid$EW(\halpha)$\mid \
\geq$ 10 \AA \ and for M3 to M5.5, $\mid$EW(\halpha)$\mid \ \geq$ 20 \AA. We
also list their multiplicity status and projected separation of companions
(when present), as determined by Brandeker et al.\ (\cite{brandeker2003}) and
Chauvin et al.\ (\cite{chauvin2002}) for 8 of the 12 members. The objects
under consideration are all T Tauri stars with spectral types between K3 and
M5.75, and at least 6 out of the 12 targets are known to have companions.
Below, we list more information on the individual targets:

\begin{description}

\item{\bf{MBM 12-1}} (RX J0255.4+2005) is a weak-line binary, with the primary
of spectral type K6 and a separation of 0.\arcsec533 (Chauvin et al.\
\cite{chauvin2002}).  X-ray observations with {\it ROSAT} showed that this object
flares, with a factor six increase in X-ray counts during the flare. Spectral
fits of the count rate suggest a coronal origin, not untypical for other
flaring WTTS (Hearty et al.\ \cite{hearty2000}).

\item{\bf{Lk\halpha \ 262}} is a CTTS, with a disc detected at 2.1~mm,
from which Itoh et al.\ (\cite{itoh2003}) estimated a disc mass of
0.05~\Msun. It is located at a distance of only 15.\arcsec3 from   
LkH$\alpha$ 263.

\item{\bf{Lk\halpha \ 263}} is a triple system, with Lk\halpha \ 262 
possible belonging to this triple to then form a quadruple system 
(Chauvin et al.\ \cite{chauvin2002}); however, it is not clear whether this
system is bound. The C component of Lk\halpha \ 263 
has spectral type M0, and was found to harbour an optically thick disc, that 
spatially-resolved observations showed to be edge-on (Jayawardhana
et al.\ \cite{rayjay2002}). These authors derived a disc mass of 0.0018~\Msun \
for this 0.7~\Msun \ C component, based on their near-IR images. Furthermore,
forbidden lines in the optical spectrum of Lk\halpha \ 263 C suggest the
presence of a jet (Jayawardhana et al.\ \cite{rayjay2002}).

\item{\bf{Lk\halpha \ 264}} is a wide binary with a separation of
9.\arcsec160.  Millimetre observations at 1.3 and 2.1~mm give a disc mass of
0.09~\Msun \ around the primary (Itoh et al.\ \cite{itoh2003}).  Emission of
molecular hydrogen at 2.1218 and 2.2233~\mic \ was also detected around the
primary. The width of these lines points to a disc origin, while further
modelling locates the NIR emitting H$_{\mathrm{2}}$ in the inner 10~AU, and
shows that the disc is seen nearly pole-on (Carmona et al.\
\cite{carmona2008}). The line strength ratio is consistent with a temperature
lower than 1500~K, and points to thermal excitation by UV photons; LkH$\alpha$
264 has a strong UV excess, so it is indeed plausible that there are enough UV
photons to excite the H$_{\mathrm{2}}$.  The total mass of the optically thin,
hot H$_{\mathrm{2}}$ in the disc of the primary is estimated to be a few lunar
masses (Carmona et al.\ \cite{carmona2008}).

\item{\bf{MBM 12-5}} (E 0255+2018) is a binary TTS with a separation of
1.\arcsec144. Following the classification by White \& Basri
(\cite{white2003}), its spectral type of K3 and \halpha \ equivalent width of
3.1~\ang \ puts it on the border between the classical and weak-line T Tauri
stars.

\item{\bf{MBM 12-6}} is a classical T Tauri star for which Brandeker et al.\
(\cite{brandeker2003}) did not find a companion within a radius of 1.\arcsec6.

\item{\bf{MBM 12-7, 9 and 11}} are all weak-line T Tauri stars, with
the latest spectral types of the whole MBM 12 sample: between M5 and M6,
corresponding to masses between 0.15 and 0.1~\Msun. We could not find any
multiplicity data for these relatively faint objects. High spatial 
resolution images are needed to establish the multiplicity status of these
sources.

\item{\bf{MBM 12-8}} is a classical TTS, with spectral type M5.5, and the
 highest \halpha \ equivalent width: 120 \AA, suggesting it is the most actively 
 accreting object of the sample. No additional information concerning its
 multiplicity status is available.

\item{\bf{MBM 12-10}} is a binary weak-line TTS with a separation of 
0.\arcsec390.

\item{\bf{S 18}} is a triple system, consisting of the primary A and, at a
projected distance of 0.\arcsec747, a tight binary companion Bab (with a 
separation of only 0.\arcsec063 - Brandeker et al.\ \cite{brandeker2003}). 
Millimetre observations at 2.1~mm give a disc mass of 0.07~\Msun \ around the
primary (Itoh et al.\ \cite{itoh2003}).
\end{description}


\subsection{{\it Spitzer} IRS spectroscopy}

The T Tauri stars in MBM 12 were observed with {\it Spitzer} as part of a
larger programme on young stellar clusters to study the evolution of
protoplanetary discs (GO proposal 20691, P.I. Bouwman).  We obtained $7.5-35$
$\mu$m low-resolution ($R = 60-120$) spectra of the MBM 12 cluster members with 
the Infrared Spectrograph (IRS, Houck et al. \ \cite{houck2004}) on-board the
{\it Spitzer Space Telescope}.  A high accuracy PCRS peak-up was executed prior 
to the spectroscopic observations to position the target within the slit.
All targets have been observed with a minimum of three observing cycles for 
redundancy.  Our spectra are based on the {\tt droopres}  products processed 
through the S15.3.0 version of the \emph{Spitzer} data pipeline.  
Partially based on the {\tt SMART} software package (Higdon et
al.\ \cite{higdon2004}), our data was further processed using spectral
extraction tools developed for the "Formation and Evolution of Planetary
Systems" (FEPS) {\it Spitzer\,} science legacy team (see also Bouwman et al.\
\cite{bouwman2008}).  The spectra were extracted using a 6.0 pixel and 5.0
pixel fixed-width aperture in the spatial dimension for the observations with
the first order of the short- ($7.5-14$ $\mu$m) and the long-wavelength
($14-35$ $\mu$m) modules, respectively. The background was subtracted using
associated pairs of imaged spectra from the two nodded positions along the
slit, also eliminating stray light contamination and
anomalous dark currents. Pixels flagged by the data pipeline as being "bad"
were replaced with a value interpolated from an 8 pixel perimeter surrounding
the errant pixel. The low-level fringing at wavelengths $>20$~$\mu$m was
removed using the {\tt irsfringe} package (Lahuis \& Boogert \cite{lahuis2003}). 
To remove any effect of pointing offsets perpendicular to the slit, we matched
orders based on the point spread function of the IRS instrument, correcting
for possible flux losses (see Swain et al. \cite{swain2008} for further
details). To remove any effect of pointing offsets parallel to the slit, we
made a flatfield correction to the nominal rsrf which was derived from
calibration measurements of standard stars where the calibrator was mapped
along the slit. The spectra are calibrated using a spectral response function
derived from multiple IRS spectra of the calibration star $\eta_{1}$~Doradus
and a MARCS stellar model provided by the {\it Spitzer\,} Science Centre. The
spectra of the calibration target have been extracted in the same way as our
science targets.  The relative errors between spectral points within one order
are dominated by the noise on each individual point and not by the
calibration. We estimate a relative flux calibration across an order of
$\approx 1$~\% and an absolute calibration error between orders/modules of
$\approx 3$~\%, which is mainly due to uncertainties in the scaling of the
MARCS model.

The targets with the latest spectral types, MBM 12-7, 8, 9 and 11,
were not detected with IRS, so we will not include them in the
rest of our analysis of the dust properties. This means that the spectral range 
for which we have IR spectroscopy is limited to K3--M5. 


\subsection{{\it Spitzer} MIPS photometry}
\label{s_mips}

\begin{table}
\begin{minipage}[t]{\columnwidth}
\caption{MIPS photometry at 24 and 70 $\mu$m, and their statistical errors
  (1$\sigma$ flux uncertainties, derived as
  described in the text). For those objects that were not
  detected, 3$\sigma$ upper limits were derived (see Sect.~\ref{s_mips}).  
  The absolute calibration uncertainties are 4 and 7\% for 24 and 70~\mic,
  respectively.
}
\label{t_mips}     
\centering                        
\renewcommand{\footnoterule}{}  
\begin{tabular}{lccc}       
\hline\hline\noalign{\smallskip}      
Object     & 24~\mic \ (1$\sigma$)   & 70~\mic \ (1$\sigma$)&\\
           & (mJy)                   & (mJy)&\\
\noalign{\smallskip}
\hline\noalign{\smallskip}
MBM 12-1   & 2.8 (0.4)            & $<$ 34.4   &\\
MBM 12-2   & 142 (0.4)            & 216\footnote{Combined flux of Lk\halpha \ 262 and 263} (12)&\\
MBM 12-3   & 50.8 (0.4)           & 216$^{a}$  (12)&\\
MBM 12-4   & 282  (0.5)           & 266 (13)       &\\
MBM 12-5   & 308  (0.5)           & 253 (12)       &\\
MBM 12-6   & 25.7 (0.4)           & $<$ 34.4   &\\
MBM 12-7   & $<$ 1.2              & $<$ 32.9   &\\
MBM 12-8   & $<$ 1.2              & $<$ 41.4   &\\
MBM 12-9   & $<$ 1.2              & $<$ 37.5   &\\
MBM 12-10  & 21   (0.4)           & $<$ 35.6   &\\
MBM 12-11  & $<$ 1.2              & $<$ 34.1   &\\
MBM 12-12  & 45.5 (0.4)           & $<$ 32.2   &\\
\noalign{\smallskip}
\hline
\end{tabular}
\end{minipage}
\end{table}

Additional infrared data was obtained using MIPS (Rieke et al.\
\cite{rieke2004}) on \emph{Spitzer} (Werner et al.\ \cite{werner2004}) in
photometry mode at 24 and 70 $\mu$m (default scale). All of our targets were
observed in February 2006, using 1 cycle of 3 sec integrations at 24 $\mu$m
and 2 - 5 cycles of 10 sec integrations at 70 $\mu$m, corresponding to
on-source intergation times of 24.1 sec and 251.6 - 629 sec at 24 and 70
$\mu$m, respectively. Our observations were processed using version S16.0.1 of
the \emph{Spitzer} Science Center (SSC) data pipeline. We created two 24
$\mu$m mosaics for each of our sources from the resulting basic calibrated
data (BCD) images of each data collection event (DCE) using the SSC's MOPEX
software (Makovitz \& Marleau \cite{mm2005}), one with a pixel scale
approximately that of the native pixel scale (2.$\arcsec$45 pixel$^{-1}$) and
another resampled to approximately one half of the native pixel scale
(1.$\arcsec$23 pixel$^{-1}$). We, similarly, created two 70 $\mu$m mosaics for
each of our sources using the filtered BCDs (which have spatial and temporal
filters applied to the data in order to remove instrumental signatures) with
outlier rejection, one with a pixel approximately that of the native pixel
scale (9.$\arcsec$9 pixel$^{-1}$) and another resampled to approximately one
quarter of the native pixel scale (2.$\arcsec$0 pixel$^{-1}$).

We used the APEX portion of MOPEX to perform aperture photometry on our
mosaicked images. APEX applies a median filter to the data to estimate the sky
background at any pixel in the image and subtracts the median-filtered image
before summing the flux in an aperture. Since the estimated backgrounds in the
majority of our fields, extrapolated from \emph{COBE (Cosmic Microwave
Background Explorer)}, was medium to high, we used a medium-sized aperture
with radius of 6$\arcsec$ at 24 $\mu$m and 16$\arcsec$ at 70 $\mu$m to reduce
the amount of background contamination in the aperture. These apertures are
not large enough to contain all of the photons from a diffraction-limited
point source; therefore, we applied scalar aperture corrections of 1.697 and
1.771 at 24 $\mu$m and 70 $\mu$m, as published on the SSC website. The 70
$\mu$m aperture correction is somewhat dependent on the colour of the source;
the aperture correction used here assumes that the measured flux has a red
power law shape, $F_{\nu}$ $\propto$ $\nu^{-2}$, because any flux detected at
70 $\mu$m is dominated by the emission from a cool, dusty disc ($T \sim$
100 K). The MIPS data handbook (version 3.2.1) states that products processed
with version S14.4 of the data pipeline have absolute calibration
uncertainties of 4 and 7\% at 24 and 70 $\mu$m, respectively.

The majority of our sources were detected with signal to noise ratios greater
than 10 at 24 $\mu$m and were not detected at 70 $\mu$m (see
Table~\ref{t_mips}). We estimate 1$\sigma$ flux uncertainties for objects that
were detected and 3$\sigma$ upper limits on the fluxes of objects that were
not detected from uncertainty mosaics produced by MOPEX (at the native pixel
resolution).  To determine the 1$\sigma$ uncertainty in a measured flux, we
take the square root of the sum of the uncertainties (in the flux in each
pixel in the aperture) in quadrature, multiplied by the aperture correction
(1.164 and 1.197 at 24 or 70 $\mu$m, respectively), centered at the expected
position of the source. Similarly, we determine the 3$\sigma$ upper limit on
the 24 and 70 $\mu$m fluxes as three times the square root of the sum of
uncertainties (in the flux in each pixel in the aperture) in quadrature,
multiplied by the aperture correction (1.164 and 1.197, respectively), in a
relatively large aperture, with radius 35$\arcsec$, centered at the expected
position of the source.

In Table~\ref{t_mips}, we list the MIPS photometry for the MBM 12 members. 
The 4 stars with the latest spectra types, MBM 12-7, 8, 9 and 11 were not 
detected at 24 nor at 70 $\mu$m. In Fig.~\ref{f_sed}, we show that the MIPS
photometry at 24~\mic \ and the fluxes from the IRS spectra agree very well.

  \begin{figure}
  \includegraphics[width=8.5cm]{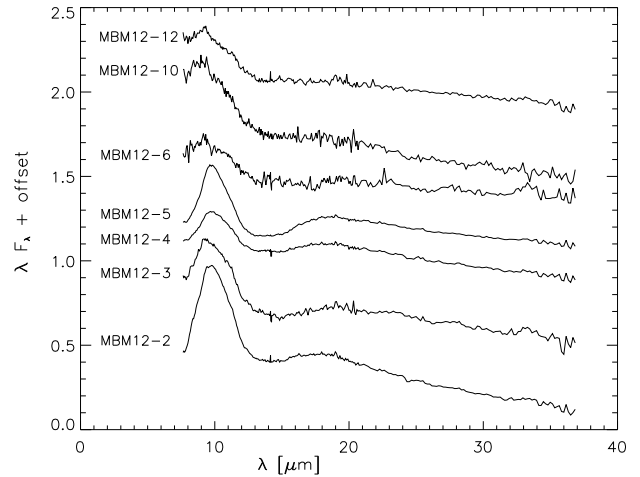}
     \caption{IRS {\it Spitzer} spectra of the seven MBM 12 sources that show
       excess emission in the infrared. The strength and shape of the emission
       features, as well the slope of the spectra varies from source to source.}
      \label{lfl}
  \end{figure}


\section{Analysis}
\label{analysis}

\subsection{Spectral energy distributions}
\label{sed}

   \begin{figure*}
   \begin{center}
   \includegraphics[width=17cm]{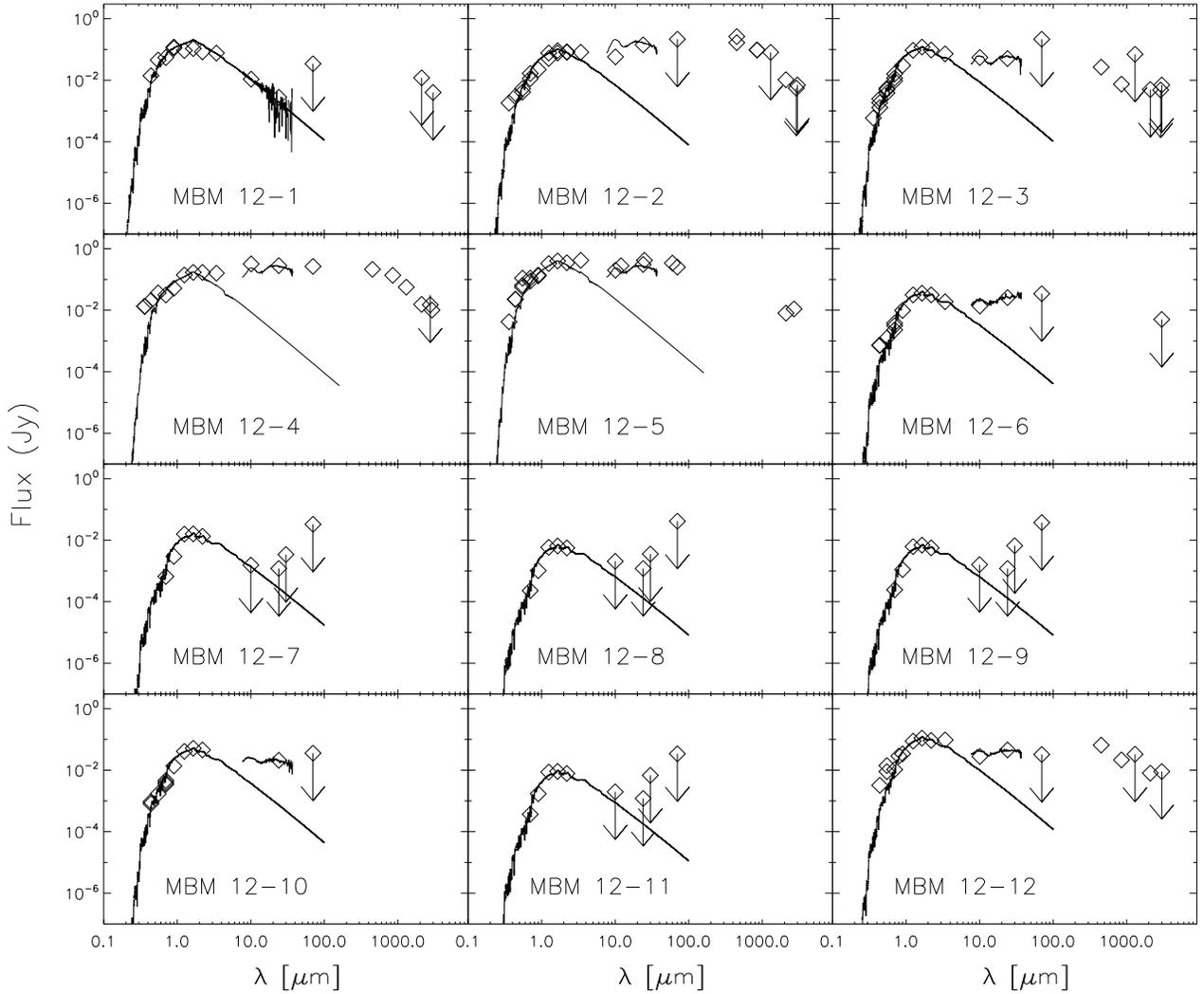}
       \caption{The spectral energy distributions of the 12 T Tauri stars in
         MBM 12. The lowest-mass stars, MBM 12-7, 8, 9 and 11 were not
         detected with {\it Spitzer}, therefore we can only show upper limits
         in the infrared for those sources.  MBM 12-1 is the only source in
         our detected sample that shows no excess above its atmosphere. MBM
         12-3, 6 and 10 appear to not have a near-IR excess, so are candidate
         transitional discs, while MBM 12-2, 4, 5 and 12 have 'normal
         accretion discs', with an excess extending into the near-IR. }
        \label{f_sed}
    \end{center}
    \end{figure*}

To complement our {\it Spitzer} observations, we collected photometry from the
literature: Jayawardhana et al.\ (\cite{rayjay2001}) and Luhman
(\cite{luhman2001}) for RI, JHK$_{s}$ and LN band photometry. The V band is
from Broeg et al.\ (\cite{broeg2006}), while the submillimetre and millimetre
photometry are from Hogerheijde et al.\ (\cite{hogerheijde2003}) and Itoh et
al.\ (\cite{itoh2003}), respectively.  In Fig.~\ref{f_sed}, we show the
spectral energy distributions (SED) of the 12 targets, together with a MARCS
stellar model (Gustafsson et al. \cite{marcs2008}) and for those sources with
\teff $>$ 4200~K with a Kurucz atmosphere model (Kurucz \cite{kurucz1994}) for
the appropriate effective temperature of the central source, as listed in
Table~\ref{tsample}. The photometry was dereddened using the A$_{\mathrm{V}}$
values derived by Luhman (\cite{luhman2001}), assuming a standard extinction
law (R$_{\mathrm{V}}$ = 3.1).

Of the 8 sources detected with {\it Spitzer}, only one object, MBM 12-1, a
K6-type weak-line TTS with a companion at 0.\arcsec533, shows no excess
emission at all: the spectrum is purely atmospheric. The other 7 sources show
an infrared excess, and also have a silicate emission feature at 10~\mic, with
varying degrees of strengths and shapes (see Fig.~\ref{lfl}). As we do not
have any data between $\sim $3.4 and 8~\mic, it is difficult to determine the
properties of the inner disc, e.g. whether there is an inner hole or not, as
witnessed by a lack of excess shortwards of the 10 micron feature. Sources MBM
12-3, 6 and 10, appear not to have a near-IR excess (see
Fig.~\ref{f_sed}); they are candidate transitional discs, where the inner disc
has already been cleared out. MBM 12-2, 4, 5 and 12 have 'full discs', as they
already have an excess in the K band. Additional data, between 3 and 8 micron,
is needed to confirm the status of the transitional candidates. An overview of
the main features observed is given in Table~\ref{t_descr}; the properties of
the dust, as derived from the solid-state features, will be discussed in the 
following section. We discuss the relation between excess luminosity and
binarity in Sect.~\ref{s_disc}.

\begin{table}
\caption{Summary of the spectral appearance of the detected targets. We
list the shape of the spectral energy distribution, where we 
distinguish between a 'full' and a 'transitional' disc, based on the 
presence or absence of excess emission at wavelengths shorter than 8 micron, 
and write 'transitional object (TO) candidate' when this classification needs
to be confirmed by additional photometry. Furthermore, we list the wavelengths
at which features of enstatite and forsterite are present. Colons 
indicate uncertain detections.}
\label{t_descr}     
\centering                        
\begin{tabular}{lccc}       
\hline\hline\noalign{\smallskip}
Object    &SED         &Enstatite &Forsterite\\
\noalign{\smallskip}
\hline\noalign{\smallskip}
MBM 12-1  &Photosphere & --       & --\\
MBM 12-2  &Full        & --       & 19, 27.8, 33~\mic \\ 
MBM 12-3  &TO candidate& 9.3:     & 11.3:, 19:, 27.8, 33~\mic \\
MBM 12-4  &Full        & --       & 11.3:, 19:, 27.8, 33~\mic \\
MBM 12-5  &Full        & --       & 19, 27.8, 33~\mic \\
MBM 12-6  &TO candidate& 9.3~\mic& 27.8, 33~\mic \\
MBM 12-10 &TO candidate& 9.3~\mic & 23.4, 27.8~\mic\\
MBM 12-12 &Full        & 9.3~\mic & 11.3, 19, 27.8, 33~\mic\\
\noalign{\smallskip}
\hline
\end{tabular}       
\end{table}


\subsection{Dust mineralogy}
\label{anaspectra}

   \begin{figure*}
   \begin{center}
\includegraphics[width=13cm]{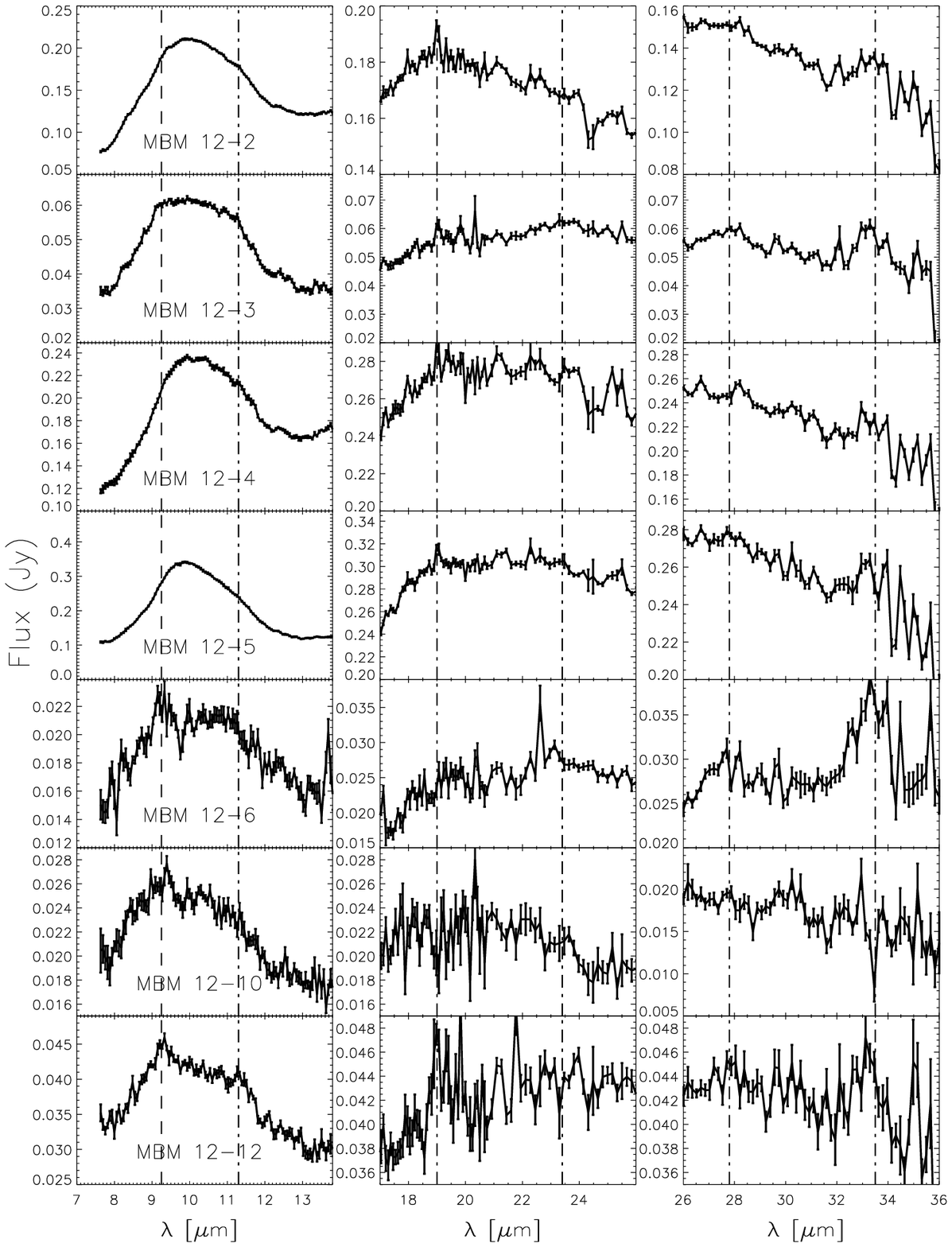}
      \caption{Zoom in on the 3 main ranges in the IRS spectra: 7-14~\mic,
        17-25~\mic \ and 26-36~\mic. We overplot the positions of the main
               crystalline features: long-dashed lines indicate enstatite,
               while dot-dashed lines indicate forsterite. }
         \label{f_ranges3}
   \end{center}
   \end{figure*}

The most frequently analysed region in the context of dust properties is the
10 micron region: it is through this window that dust characteristics like
composition (e.g. olivines, Mg$_{{\rm 2x}}$Fe$_{\rm 2(1-x)}$SiO$_{\rm 4}$ and
pyroxenes, Mg$_{\rm x}$Fe$_{\rm{ (1-x)}}$SiO$_{\rm 3}$, or Mg/Fe content),
structure (crystalline vs. amorphous) and grain sizes can be derived from
their solid-state features. The 10~\mic \ region is only sensitive to dust
grains up to sizes of a few microns, as the feature flattens out when the dust
grains attain sizes similar to the wavelength, making them essentially
invisible. Furthermore, it is important to realise that the dust causing this
feature traces only a small fraction of the disc material, namely those dust
grains that are located in the optically thin disc atmosphere, while the bulk
of the dust mass is located in the disc midplane. In addition, as the dust
temperature needs to be high enough (150-450 K) in order to radiate at
10~\mic, the radial location of the dust observed is also limited: for a
typical star in our sample, this is around 1~AU. At longer wavelengths,
features of crystalline dust (forsterite and enstatite) witness another
important part of dust processing in the protoplanetary disc.

In Fig.~\ref{f_ranges3}, we zoom in on the three most interesting ranges that
are observed with IRS, and also indicate the location of the most important
features. Apart from the broad feature of amorphous silicate at 9.7 (present
in all 7 stars with a detected excess) and 18~\mic, features of enstatite
(MgSiO$_{\rm 3}$) and forsterite (Mg$_{\rm 2}$SiO$_{\rm 4}$) are observed. We
do not see evidence for carbonaceous dust, such as features from polycyclic
aromatic hydrocarbons (PAHs). This is perhaps not surprising, given that PAHs
are transiently excited by UV photons and the central sources have low
temperatures, although some sources do show a rather large UV excess (e.g. MBM
12-2, 4 and 12).

  \begin{figure*}
  \begin{center}
  \includegraphics[width=9cm,angle=90]{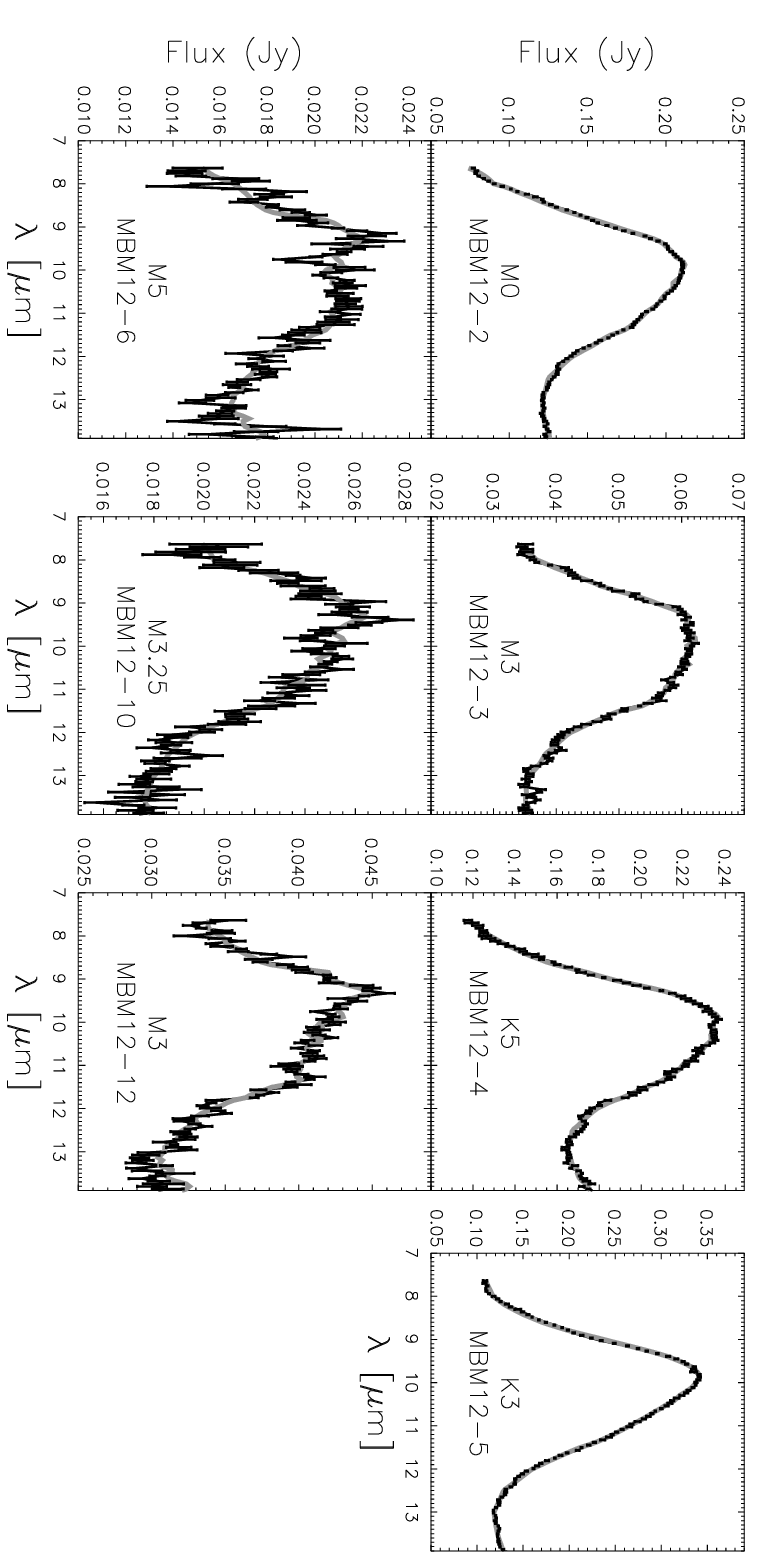}
     \caption{The TLTD fits of the 10 micron silicate feature (thick, solid
     line), and the IRS spectra with their noise (black points with error
     bars). Notice the different shapes that are present in this small
     sample: MBM 12-5 has a triangular shape, pointing to small, amorphous 
     silicate grains, while MBM 12-6 has a much broader feature, indicating
     larger and crystalline grains. The narrow emission feature at 9.3~\mic \ 
     is caused by crystalline enstatite.}
        \label{fit10}
  \end{center}
  \end{figure*}

\begin{table}
\caption{Overview of dust species used in our fitting routines. For each
     component, we specify its lattice structure, chemical composition, shape
 and reference to the laboratory measurements of the optical constants. For
 the homogeneous spheres, we used Mie theory to calculate the opacities. For
 the inhomogeneous spheres, we used the distribution of hollow spheres (Min et
 al.\ \cite{min2005}), to simulate grain shapes deviating from perfect
 symmetry.}
\begin{tabular}{llll}
\hline\hline
\noalign{\smallskip}
Species & State & Chemical & Shape\\
        &       & formula  &       \\
\noalign{\smallskip}
\hline
\noalign{\smallskip}
Amorphous silicate$^{1}$  & A  &  MgFeSiO$_{4}$       & Homogeneous \\
(Olivine stoichiometry)& &                      &     Sphere          \\
Amorphous silicate$^{1}$  & A  &  MgFeSi$_{2}$O$_{6}$ & Homogeneous \\
(Pyroxene stoichiometry)&                      &             & Sphere  \\
Forsterite$^{2}$   & C  &  Mg$_{2}$SiO$_{4}$   & Hollow  Sphere    \\
Clino Enstatite$^{3}$& C  &  MgSiO$_{3}$         & Hollow  Sphere    \\
Silica$^{4}$     & A  &  SiO$_{2}$           & Hollow  Sphere    \\
\noalign{\smallskip}
\hline
\noalign{\smallskip}
\multicolumn{4}{l}{References: (1) Dorschner et al. \ \cite{dorschner1995};
  (2) Servoin et al.\ \cite{servoin1973}; }\\ 
\multicolumn{4}{l}{(3) J\"ager et al.\ \cite{jaeger1998}; (4) Henning et al. \
  \cite{henning1997}}\\
\noalign{\smallskip}
\hline
\end{tabular}
\label{t_species}
\end{table}

In order to analyse the composition of the dust in the disc atmosphere, the
radiation of which dominates the IRS spectrum, we use the two-layer
temperature distribution (TLTD) spectral decomposition routines described in
Juh\'asz et al.\ (\cite{juhasz2008}).  This method uses a multi-component
continuum (star, inner rim, disc midplane), assuming that the region where the
observed radiation originates (both optically thin and thick) has a
distribution of temperatures. In this fitting method, the observed
flux-density at a given frequency is given by

\begin{eqnarray}
F_\nu = F_{\nu, {\rm cont}} & + & \sum_{i=1}^N\sum_{j=1}^MD_{i,j}\kappa_{i,j}
\int_{{T_{a, \rm in}}}^{{T_{a, \rm out}}}\frac{2\pi}{d^2}B_\nu(T){T}^{\frac{2-qa}{
qa}}dT
\label{eq:1}
\end{eqnarray}

where $N$ and $M$ are the number of dust species and of grain
sizes used, respectively.  $\kappa_{i,j}$ is the mass absorption coefficient of
the dust species $i$ and grain size $j$. $B_\nu(T)$ is the Planck-function,
$qa$ is the power exponent of the temperature distribution and $d $ is the
distance to the source. The subscript '$a$' in the integration boundaries refers
to the disc atmosphere. The continuum emission ($F_{\nu, {\rm cont}}$) is
given by

\begin{eqnarray}
F_{\nu, {\rm cont}} = D_0 \frac{\pi R_\star^2}{d^2} B_\nu(T_\star)&+&
D1\int_{{T_{r, \rm in}}}^{{T_{r, \rm out}}}\frac{2\pi}{d^2}B_\nu(T){T}^{\frac{2-qr}{qr}}dT \\
&+& D2\int_{{T_{m, \rm in}}}^{{T_{m, \rm out}}}\frac{2\pi}{d^2}B_\nu(T){T}^{\frac{2
-qm}{qm}}dT.
\label{eq:2}
\end{eqnarray}

The first term on the right hand side describes the emission of the star,
while the second and third term describe the radiation of the inner rim
(subscript '$r$') and the disc midplane (subscript '$m$'), respectively. The
meaning of the different parameters are summarised in Table~\ref{tparam} in
the Appendix. For each component (disc atmosphere, inner rim, midplane), the
highest temperature is fitted to obtain $T_{a/r/m,\ \rm in}$, while the lowest
temperature, $T_{a/r/m,\ \rm out}$, is calculated requiring that the annulus
with that temperature contributes more than 0.1\% to the total flux.

For our fit, we used five dust species that are commonly found in the discs of
young stars, in three grain sizes (0.1, 1.5 and 6\,$\mu$m).  We
applied the theory of distribution of hollow spheres for the crystalline dust
(Min et al.\ \cite{min2005}) and the classical Mie theory for spherical
particles for the amorphous dust, to derive the mass absorption coefficients
from the optical constants. The list of the dust species, the origin of the
optical constants and the grain model used are presented in
Table~\ref{t_species}. Furthermore, since the dust grains radiating at
8-13~\mic\ and 20-30~\mic \ are at different temperatures, hence at different
radial distances from the star, we splitted the wavelength region in two parts
before fitting. This way, we can take into account that different species can
contribute in different amounts to the spectrum, if there is a radial gradient
in their abundances and/or properties. We chose the 1) the 7-17~\mic \ range
for the shorter wavelength region, which Juh\'asz et al.\ (\cite{juhasz2008})
showed to be the optimal wavelength range to balance the presence of spectral
features and radial changes, and 2) the 17-37~\mic \ range for the longer
wavelength region, which includes important crystalline features; both regions
were separatedly fitted, and we will refer to them as the 'shorter' and
'longer' wavelength regions.

As the IRS spectra do contain noise, we repeated the fit of each object
100 times, every time adding random Gaussian noise on top of the measured
spectrum. The final composition is derived as the average of those 100 fits,
while the errors are derived from the standard deviation, taking into account
the positive and negative directions of the variations. Therefore, the errors
we give describe the S/N and measurement errors in the spectra. We
emphasize that we cannot give error estimates that are related to species that
are not included in our fits, nor of different shapes or composition of the
grains. Those errors are subject of another study (A. Juh\'asz et al. 2009,
{\it in preparation}).

With the fits, we determine the source of the solid-state features that are
observed in the spectra. Since several species (e.g. carbon) are featureless
within our wavelength range (but can contribute to the continuum), they remain
undetected in our fits; so it is important to realise that we can only discuss
the 'visible dust', that has features in the range 7 to 37 micron. In
Fig~\ref{fit10}, we show the best fit of the 7 to 17 micron region, and a
summary of the derived dust properties for both the shorter and longer
wavelength regions is given in Table~\ref{tfit_summ}. The detailed results of
the fits in both regions, in terms of mass fractions of the relevant species
and their sizes, are given in Table~\ref{tfit} in the Appendix.

\begin{table}
\caption{Distilled fitting results: given are mass-averaged
       size of the amorphous and the crystalline silicates,
       $<a>_{\mathrm{am. sil.}}$ and $<a>_{\mathrm{cryst. sil.}}$,
       respectively, mass fraction of the crystalline grains,
       f$_{\mathrm{cryst.}}$ (ratio of crystalline silicate to total silicate
       mass), and the mass ratio of forsterite to enstatite, as derived for
       both the 7-17 and 17-37~\mic \ regions. }
\label{tfit_summ}     
\centering                        
\begin{tabular}{lcccc}       
\hline\hline\noalign{\smallskip}
Object&$<a>_{\mathrm{am. sil.}}$&$<a>_{\mathrm{cryst. sil.}}$&f$_{\mathrm{cryst.}}$&forst/enst\\
      &(\mic)                   &(\mic)               &  (\%)    &             \\
\noalign{\smallskip}
\hline\noalign{\smallskip}
\multicolumn{5}{c}{7-17~\mic \ region}\\
\noalign{\smallskip}
MBM 12-2             & 4.1 ($^{+0.1}_{-0.1}$)    & 5.2 ($^{+0.1}_{-0.1}$)    & 10.7 ($^{+0.7}_{-0.8}$)   & 0.1 ($^{+0.0}_{-0.0}$)    \\
MBM 12-3             & 4.0 ($^{+0.4}_{-0.6}$)    & 1.7 ($^{+1.7}_{-1.1}$)    & 5.8 ($^{+2.1}_{-1.6}$)    & 0.4 ($^{+0.1}_{-0.2}$)    \\
MBM 12-4             & 0.4 ($^{+0.1}_{-0.1}$)    & 4.1 ($^{+0.5}_{-0.6}$)    & 16.3 ($^{+2.4}_{-2.6}$)   & 0.1 ($^{+0.0}_{-0.0}$)    \\
MBM 12-5             & 2.2 ($^{+0.1}_{-0.0}$)    & 4.1 ($^{+1.0}_{-1.6}$)    & 2.9 ($^{+1.2}_{-1.1}$)    & 0.3 ($^{+0.3}_{-0.1}$)    \\
MBM 12-6             & 3.9 ($^{+2.0}_{-3.1}$)    & 5.3 ($^{+0.2}_{-0.2}$)    & 46.5 ($^{+24.9}_{-16.6}$) & 0.0 ($^{+0.0}_{-0.0}$)    \\
MBM 12-10            & 5.3 ($^{+0.1}_{-0.1}$)    & 1.9 ($^{+2.1}_{-0.9}$)    & 4.6 ($^{+4.6}_{-1.7}$)    & 0.8 ($^{+2.2}_{-0.5}$)    \\
MBM 12-12            & 6.0 ($^{+0.0}_{-0.1}$)    & 5.5 ($^{+0.1}_{-0.1}$)    & 32.8 ($^{+2.9}_{-2.8}$)   & 0.1 ($^{+0.1}_{-0.0}$)    \\
\noalign{\smallskip}
\hline\noalign{\smallskip}
\multicolumn{5}{c}{17-37~\mic \ region}\\
\noalign{\smallskip}
MBM 12-2             & 0.1 ($^{+0.1}_{-0.0}$)    & 0.8 ($^{+0.0}_{-0.0}$)    & 1.8 ($^{+0.1}_{-0.1}$)    & 1.1 ($^{+0.1}_{-0.1}$)    \\
MBM 12-3             & 1.2 ($^{+0.1}_{-0.1}$)    & 0.4 ($^{+0.3}_{-0.2}$)    & 4.8 ($^{+0.4}_{-0.4}$)    & 1.2 ($^{+0.1}_{-0.1}$)    \\
MBM 12-4             & 0.2 ($^{+0.2}_{-0.1}$)    & 1.6 ($^{+0.8}_{-0.6}$)    & 2.8 ($^{+0.8}_{-0.5}$)    & 1.0 ($^{+0.3}_{-0.2}$)    \\
MBM 12-5             & 0.7 ($^{+0.2}_{-0.1}$)    & 4.0 ($^{+0.5}_{-0.8}$)    & 3.5 ($^{+1.0}_{-0.9}$)    & 0.9 ($^{+0.7}_{-0.3}$)    \\
MBM 12-6             & 1.5 ($^{+0.4}_{-0.2}$)    & 0.1 ($^{+0.0}_{-0.0}$)    & 7.1 ($^{+1.5}_{-1.3}$)    & --                        \\
MBM 12-10            & 0.4 ($^{+0.9}_{-0.3}$)    & --                        & 2.9 ($^{+3.1}_{-1.5}$)    & --                        \\
MBM 12-12            & 0.6 ($^{+0.3}_{-0.2}$)    & 0.8 ($^{+0.2}_{-0.2}$)    & 3.3 ($^{+1.1}_{-0.7}$)    & 0.9 ($^{+0.2}_{-0.1}$)    \\
\noalign{\smallskip}
\hline
\end{tabular}       
\end{table}


\section{Discussion: 2 Myr-old T Tauri discs}
\label{discussion}

\subsection{Grain growth}

  \begin{figure}
  \includegraphics[width=9cm]{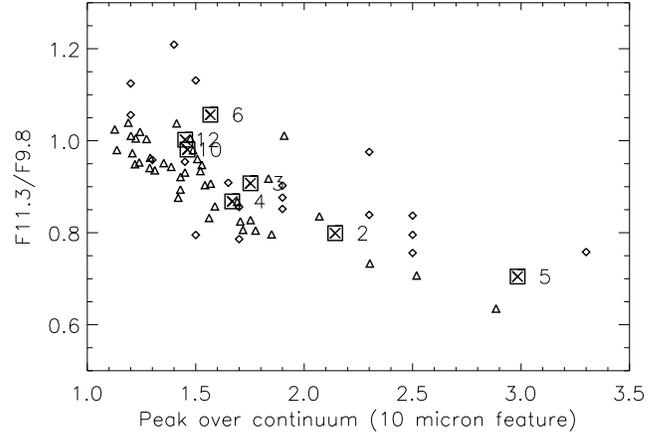}
     \caption{Relation between the shape of the 10~\mic \ feature (flux ratio 
     11.3 over 9.8~\mic) and the feature strength (peak over continuum).  Our
     objects (squared symbols, numbered as listed in Table~\ref{tsample}),
     follow the trend observed for TTS in Tr 37 (4~Myr) and NGC 7160 (12~Myr),
     shown by diamonds (Sicilia-Aguilar et al.\ \cite{aurora2007}) and TTS in
     Taurus (2~Myr), shown by triangles (Pascucci et al.\ \cite{pascucci2008}): the
     flux ratio is correlated with the feature strength. }
        \label{fratiopeak}
  \end{figure}

A first thing to notice, when looking at the 10 micron silicate feature, is
that the strength of the emission varies. Van Boekel et al.\
(\cite{boekel2003}) showed that the shape and the strength of the 10~\mic \
feature in Herbig Ae/Be stars are related, and demonstrated it to be evidence
for grain growth: a strong and triangular 10~$\mu$m feature is typical of
small (submicron-sized) grains, whereas a weaker and broader feature indicates
the presence of larger sized grains (up to a few microns). In
Fig.~\ref{fratiopeak}, we probed this relation for our sample of T Tauri
stars, and also found evidence for grain growth in the protoplanetary discs of
the MBM 12 members.

  \begin{figure}
  \includegraphics[width=9cm]{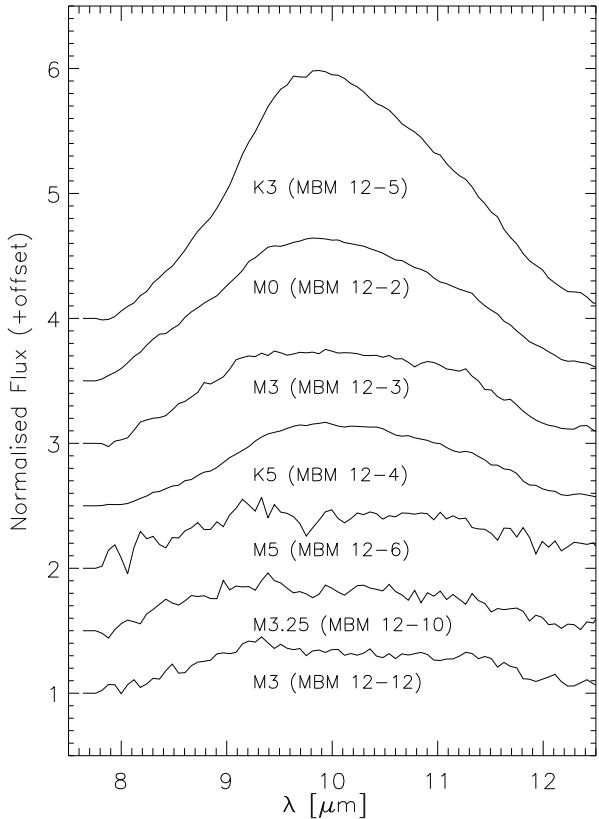}
     \caption{Continuum normalised flux in the 10~\mic \ region. The objects 
      with the latest spectral types tend to have the weakest feature.}
        \label{f10ontop}
  \end{figure}

\begin{figure}
\includegraphics[width=9cm]{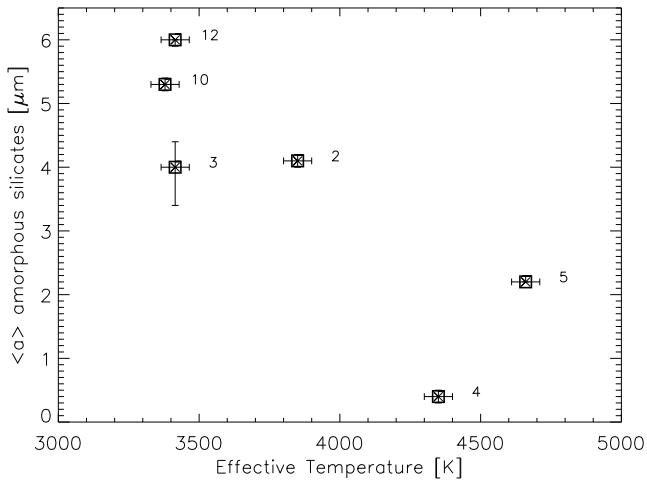}
    \caption{Mass-averaged size of the amorphous silicate grains in the warmer
      region, versus effective temperature of the central star. The cooler the
         star (the later the spectral type), the larger the dust grains tend to be.}
         \label{amsize_sptype}
   \end{figure}

Kessler-Silacci et al.\ (\cite{kessler2006,kessler2007}) and Apai et
 al. (\cite{apai2005}) noticed that later type TTS have weaker 10 micron
 features than earlier type TTS, and related this to the location of the dust
 causing this feature. This relation was confirmed by Sicilia-Aguilar et al.\
 (\cite{aurora2007}), who found that the presence of a very weak feature was 3
 times more frequent for M-type stars than for earlier-type stars. The
 temperature of the silicates causing the feature is $\sim$ 300~K, a
 temperature that is reached more inwards for lower luminosity sources than
 for higher luminosity sources: under the assumption of a similar disc
 structure for all spectral types discussed here, an M0-type star reaches
 300~K between 0.7 and 1.5~AU, while it is between 1.2 and 3~AU for a K5-type
 star, a factor 2 difference in distance.  Furthermore, the density
 distribution has a radial dependence, as it decreases with increasing radius,
 and grain growth occurs faster in more dense environments. This implies that
 grain growth will be faster in the more inward regions, so that lower
 luminosity sources will {\it appear} to have larger grains, as their 10
 micron feature originates from a denser region, in which growth naturally
 occurs more rapidly. This is also seen in our spectra: in
 Fig.~\ref{f10ontop}, we plot the continuum-normalised spectra, and also list
 their spectral types. With the exception of MBM 12-4, the fainter objects
 indeed have weaker features. This is further quantified through our fitting
 with the TLTD method: in Fig.~\ref{amsize_sptype}, we show the relation
 between the mass-averaged grains size in the 10 micron region and the
 effective temperature of the central star. There is a clear trend for cooler
 stars to have larger grain sizes.

Of course, not all discs need to have the same structure, and the degree of
flaring can also play an important role in this context. However, to study
these effects, detailed radiative transfer modelling of each source is needed,
which is beyond the scope of this paper.

\subsection{Crystallisation}

    \begin{figure}
 \includegraphics[width=9cm]{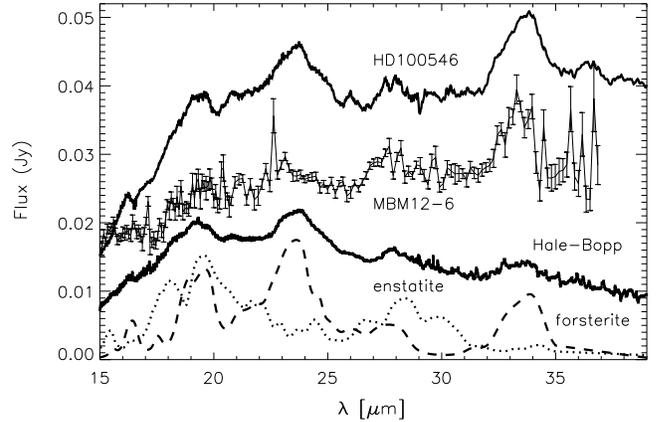}
     \caption{MBM 12-6, host of a large amount of crystalline silicates,
       witnessed by the agreement in peak positions between the spectrum and
       the features of the mass absorption coefficients of enstatite (J\"ager
       et al.\ \cite{jaeger1998}; dotted line) and forsterite (Servoin et al.\
       \cite{servoin1973}; dashed line).  For comparison, we also show the
       Herbig Be star HD100546 (Malfait et al.\ \cite{malfait1998}), and the
       solar-system comet Hale-Bopp (Crovisier et al.\ \cite{crovisier1997}),
       both host of large amounts of crystalline dust. The fluxes of
       HD~100546 and Hale-Bopp are scaled to match the flux of MBM 12-6.}
         \label{lu6vs}
   \end{figure}

\begin{figure}
    \includegraphics[width=9cm]{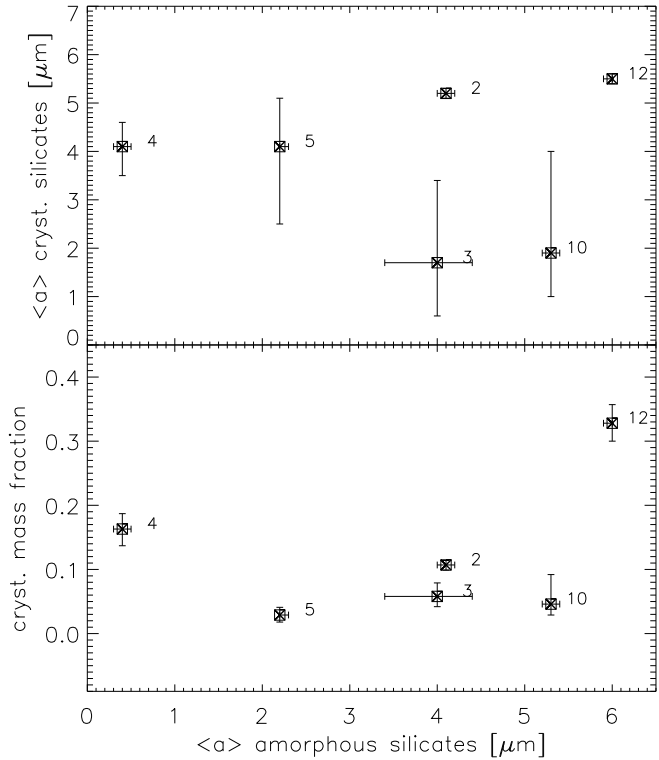}
     \caption{Relation between the properties of the crystalline silicates 
     and the size of the amorphous silicates in the 7 to 17 micron
     region. Upper panel: mass-averaged size of the crystalline silicates
     versus mass-averaged size of the amorphous silicates. The two appear
     unrelated. Lower panel: crystalline mass fraction versus mass-averaged
     size of the amorphous silicates. Also here we do not see a correlation.}
         \label{f_cryst_amorf}
   \end{figure}

\begin{figure}
   \includegraphics[width=9cm]{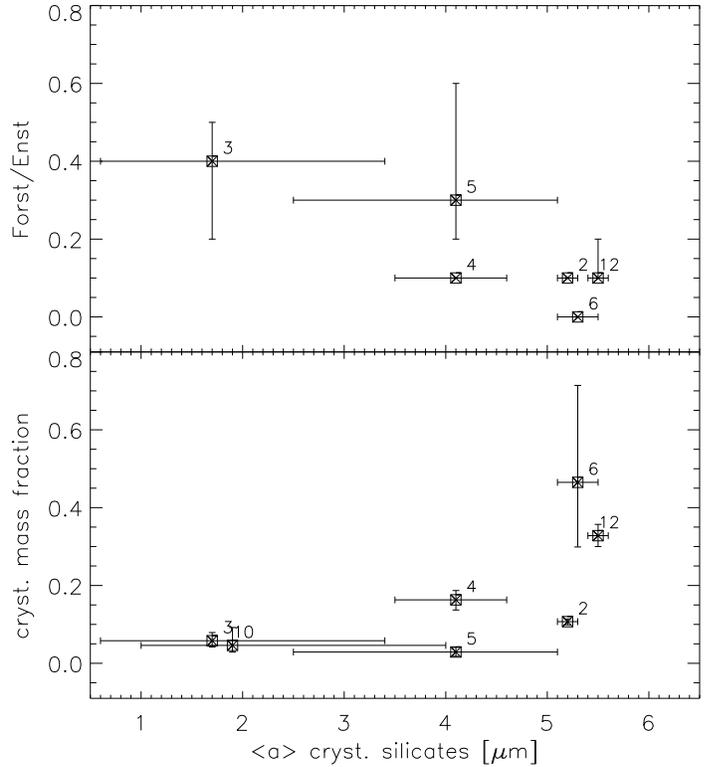}
      \caption{Properties of the the crystalline silicates, derived from the 7
        to 17 micron region. Upper panel: the forsterite over enstatite ratio
        in function of the size of the crystalline silicates. There is a trend
        for larger grains to have a smaller forst/enst ratio. Lower panel:
        crystallinity in function of crystalline grain size . The crystalline
        fraction only exceeds 20\% when the crystalline grains are larger than
        5 micron.  }
         \label{f_cryst}
   \end{figure}

The dust that is initially incorporated into the protoplanetary disc is
largely amorphous, as it comes from the ISM, for which an upper limit of $\sim
$2\% in mass of the crystalline grains was determined using spherical grains
(Kemper et al.\ \cite{kemper2004}).  Min et al.\ (\cite{min2007}) further
improved this number using grains with irregular shapes and derived a
crystallinity of only $\sim$1\%.  However, the amorphous dust may become
crystalline through thermal annealing (e.g. Fabian et al.\ \cite{fabian2000})
or shock heating (e.g. Scott \& Krot \cite{scott2005}). We derived the
crystalline mass fraction in the shorter wavelength region, and found a large
degree of variation: between 2.9$^{+1.2}_{-1.1}$~\% for MBM 12-5 and
47$^{+25}_{-17}$~\% for MBM 12-6.  The crystalline fraction of MBM 12-6
is not very well determined, therefore we use the lower limit of 30~\% that
was found for this object. The mass fraction of the crystalline silicates is
not related to the spectral type. Furthermore, we do not find a correlation
between crystallinity and the size of the amorphous grains; also the size of
the crystalline grains appears to be uncorrelated with amorphous grain size
(see Fig.~\ref{f_cryst_amorf}).
However, we do note that the sources with the largest crystalline mass
fractions (MBM 12-6 and MBM 12-12) also have the largest grains size, both for
the amorphous as for the crystalline grains (see also Fig.~\ref{f_cryst}).  In
the longer wavelength region, we find the variation in crystallinity to be
less, and the crystalline mass fraction smaller: between
1.8$^{+0.1}_{-0.1}$~\% for MBM 12-2 and 7.1$^{+1.5}_{-1.3}$~\% for MBM
12-6. This difference in crystallinity between both regions for the whole
sample could be related to the presence of a large amount of small amorphous
silicate grains in the cooler disc region (average size for the sample is
0.7~\mic), while in the warmer disc region the amorphous silicate grains are
much larger (average size for the sample is 3.7~\mic). Larger grains will
produce less detectable features, so that the crystalline grains that still
show features will appear more abundant.

In Fig.~\ref{lu6vs}, we show the 15 to 37~\mic \ spectrum of MBM 12-6, the
object with the largest fraction of crystalline grains. Its IR spectral 
appearance is similar to that of HD~100546, a Herbig B9e star with highly 
evolved dust, which is remarkable, given that their effective temperatures are 
very different: 11000~K for the B9 star, and only 3200~K for the M5 type
star. Still, the dust around both objects is dominated by crystalline grains,
indicating that the temperature of the central star does not play an important
role in the crystallisation process. The spectrum of MBM 12-6 is also similar
to that of solar-system comet Hale-Bopp, so that similar dust processing
mechanisms must exist in both our solar system and the disc of MBM 12-6.

In the dust model by Gail (\cite{gail1998}), the composition of the
dust is derived, based on condensation sequences and chemical equilibrium 
considerations. He predicts that - assuming crystalline silicates form as 
high temperature gas phase condensates - of the crystalline silicates, 
enstatite will be the dominant constituent, while forsterite is only present 
in a small region close to the star.  More recent simulations show that radial
mixing within the disc is capable of transporting a substantial amount of 
crystalline material from the inner disc towards more colder, outward regions 
(Keller \& Gail \cite{keller2004}).

To better understand the crystallisation process occurring in the T Tauri
discs, we look at the forsterite to enstatite ratio. In Table~\ref{tfit_summ}, 
we list the forsterite to enstatite mass ratio for both regions. In the warmer 
region, the smallest ratio is found for those sources with the largest grains
(see Fig.~\ref{f_cryst}), 
implying that enstatite is more abundant (relative to forsterite) when larger 
grains are present. We further compare the relative mass fractions of
forsterite and enstatite in both regions: the average ratio for
the sample is 0.3 in the warmer region, and 0.9 in the cooler region.

It is thus clear that there is a spatial gradient in the forsterite to
enstatite ratio, with forsterite dominating the cooler regions, and enstatite
more abundant in the warmer, inner regions. Such a gradient was already
found in a few earlier studies, e.g. Bouwman et al.\ (\cite{bouwman2008}). 
This is in contrast with the predictions by Gail (\cite{gail1998,gail2004}),
suggesting that the chemical equilibrium conditions needed for the forsterite
to enstatite conversion are not reached inside these discs, and that the
crystallisation process must be different, even when including radial
mixing. It is interesting to consider that enstatite is more abundant in the
warmer (inner) regions, and that we also find a larger enstatite abundance in
those sources that have larger grains (see Fig.~\ref{f_cryst}). These
observations suggest that enstatite forms easier in regions of higher density,
where also grain growth is more abundant.  Bouwman et al.\
(\cite{bouwman2008}) discuss the formation of enstatite and forsterite in
detail, and link the formation of enstatite in the inner region with the
conditions that prevail there: due to the higher density and temperature, it
takes the dust grains longer to cool down so that, potentially, equilibrium
conditions can be reached - what is more unlikely at larger radial distances,
where both the density and the temperature are lower.

\subsection{Disc properties}
\label{s_disc}

\subsubsection{Disc fraction}

  \begin{figure}
   \includegraphics[width=9cm]{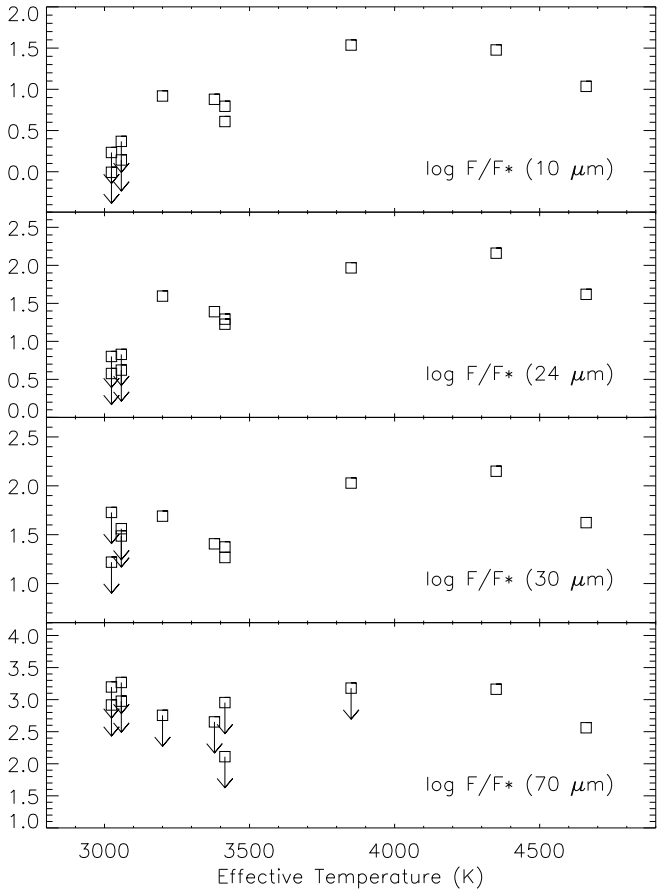}
      \caption{The logarithm of the IR excess at selected wavelengths, plotted
        in function of the effective temperature of the central star. Upper
      limits are indicated with arrows. The warmer the star, the larger the
      infrared excess - apart for the warmest object, MBM 12-5, with a 
      temperature of 4660~K.}
         \label{f_excess}
   \end{figure}

We obtained {\it Spitzer} data for the complete sample of MBM 12 members; 4 of
them, however, could not be detected.  The remaining 8 objects (spectral type
range K3 to M5) have a disc fraction rate of 7 out of 8, or nearly 90\%. This 
is very high, when compared to a similar spectral type range in other star
forming regions: Damjanov et al. (\cite{damjanov2007}) derive a disc fraction
of 47\% for M0-M4 and 55\% for K3-K8 type stars in the 2 Myr-old Chamaeleon I
and 63\% 
region; however, Flaherty and Muzzerole (\cite{flaherty2008}) derive a disc
fraction between 65 and 81\% for the 2 Myr old clusters NGC 2068 and NGC2071.
The high disc fraction rate found for MBM 12 may (partly) be attributed to the
absence of close companions to the disc bearing stars:
there are no companions found at a projected distance smaller than 0.\arcsec39
(or 70 AU at the distance of MBM 12).

Of the 7 T Tauri stars with a disc, 3 are candidate transitional objects,
which is a high rate for a region with such a young age (2 Myr), pointing to
fast inner disc dispersal. However, due to the lack of data between 3.5 and
8~\mic, we cannot further discuss the differences in inner disc structure;
additional observations could shed more light on this topic. For the four
non-detected members, we derived upper limits, to put some constraints on
their disc properties. In all 4 cases, the upper limit at 10 micron does not
allow for an IR excess to be present, suggesting that, if these objects do
have a disc - what cannot be excluded from the upper limits derived at longer
wavelengths - these discs must have inner holes.  In Table~\ref{t_excess}, we
list the excess luminosity over the stellar photosphere; the data are
visualised in Fig.~\ref{f_excess}, where we plot the excesses versus the
effective temperature. Those sources that have the largest excesses, also have
the highest temperatures, although there is no linear relation between both
quantities.

\begin{table}
\begin{minipage}[t]{\columnwidth}
\caption{The logarithm of the IR excess over the photosphere at 10, 24, 30
  and 70~\mic. Upper limits are indicated with '$<$'. }
\label{t_excess}     
\centering                        
\renewcommand{\footnoterule}{}  
\begin{tabular}{lcccc}       
\hline\hline\noalign{\smallskip}
           & $\log$  & $\log$  & $\log$  & $\log$  \\
Object     & $(F_{10}/F_{*,10})$ & $ (F_{24}/F_{*,24})$ & $ (F_{30}/F_{*,30})$ & $ (F_{70}/F_{*,70})$ \\
\noalign{\smallskip}
\hline\noalign{\smallskip}
MBM 12-1   & 0.0  & 0.0 & 0.0 & 2.1\\
MBM 12-2   & 1.5  & 2.0 & 2.0 & $<$ 3.2\footnote{Combined flux of Lk\halpha \ 262 and 263} \\
MBM 12-3   & 0.8  & 1.3 & 1.4 & $<$ 3.0$^{a}$ \\
MBM 12-4   & 1.5  & 2.2 & 2.1 & 3.2\\
MBM 12-5   & 1.0  & 1.6 & 1.6 & 2.6\\
MBM 12-6   & 0.9  & 1.6 & 1.7 & $<$ 2.8\\
MBM 12-7   & $<$ 0 & $<$ 0.6 & $<$ 1.2 & $<$ 2.9\\
MBM 12-8   & $<$ 0.4  & $<$ 0.8 & $<$ 1.5 & $<$ 3.3\\
MBM 12-9   & $<$ 0.2  & $<$ 0.8 & $<$ 1.7 & $<$ 3.2\\
MBM 12-10  & 0.9  & 1.4 & 1.4 & $<$ 2.7\\
MBM 12-11  & $<$ 0.1  & $<$ 0.6 & $<$ 1.6 & $<$ 3.0\\
MBM 12-12  & 0.6  & 1.2 & 1.3 & $<$ 2.1\\
\noalign{\smallskip}
\hline
\end{tabular}
\end{minipage}
\end{table}

\subsubsection{The influence of companions}
The radiation of cold, outer discs was detected by Hogerheijde et al.\
(\cite{hogerheijde2003}) at 450 and 850~$\mu$m for 4 stars in MBM 12:
Lk\halpha \ 262 (MBM 12-2), Lk\halpha \ 263 ABC (MBM 12-3), Lk\halpha \ 264 A
(MBM 12-4) and S18 ABab (MBM 12-12). These authors derived disc masses between
0.005 and 0.23 $\times 10^{-2}$ M$_{\sun}$ for these 4 sources, and noticed a
relation between the cold disc masses (as derived from the submm photometry)
and the distance of a companion: the 2 stars with separations of 100-200~AU
have much smaller disc masses than the 2 stars with separations of
2000-4000~AU.  Artymowicz \& Lubow (\cite{artym1994}) discussed disc
truncation by companions, and showed that the tidal limit for a
disc around a member of a binary is $\sim$ 0.4 times the separation. For 
objects with a not too wide companion, relative to their disc size, this means 
that the companion will effectively truncate their disc. Jensen et
al.\ (\cite{jensen1996}) showed with millimetre observations that indeed,
objects with separations less than $\sim$ 100~AU have smaller disc masses,
when compared to their single counterparts, or to those objects which have
only wider companions.

We now want to see what the influence of companions is on the 'middle' region
of the disc, that can be observed in the mid-IR. Therefore, we calculated the
IR excess at 30 micron for 2 groups we defined in our sample: 1) the
'binaries', including binaries with projected separations smaller than 400 AU
(MBM 12-3, 5, 10 and 12), and 2) the 'singles', including single stars
and binaries with larger projected separations (MBM 12-2, 4 and 6). This gives
average excess values of $F_{30}/F_{*,30} \sim$ 25 for the binaries, and
$F_{30}/F_{*,30} \sim$ 80 for the singles, indicating that the discs
around the binaries are indeed influenced by their companions, as their
excess is a factor 3 smaller than that of the single stars.

Bouwman et al.\ (\cite{bouwman2006}) studied a sample of T Tauri stars in the
8 Myr-old cluster $\eta$ Cha, and found that only 1 out of 6 known or
suspected binaries (all with projected separations $\leq 20$ AU) retain a
protoplanetary disc. Conversely, 7 out of 9 stars believed to be single have
discs, suggesting that binary discs have shorter lifetimes. Our MBM 12 sample 
does not include such close companions around the disc-bearing stars, but
nevertheless shows a dependence of its dust characteristics on the presence of
a companion. It is interesting to see that the two groups (binaries and
singles) already differ in disc properties at such a young age, even when
there are no very close companions involved.

\subsubsection{Disc flaring}

  \begin{figure}
\includegraphics[width=9cm]{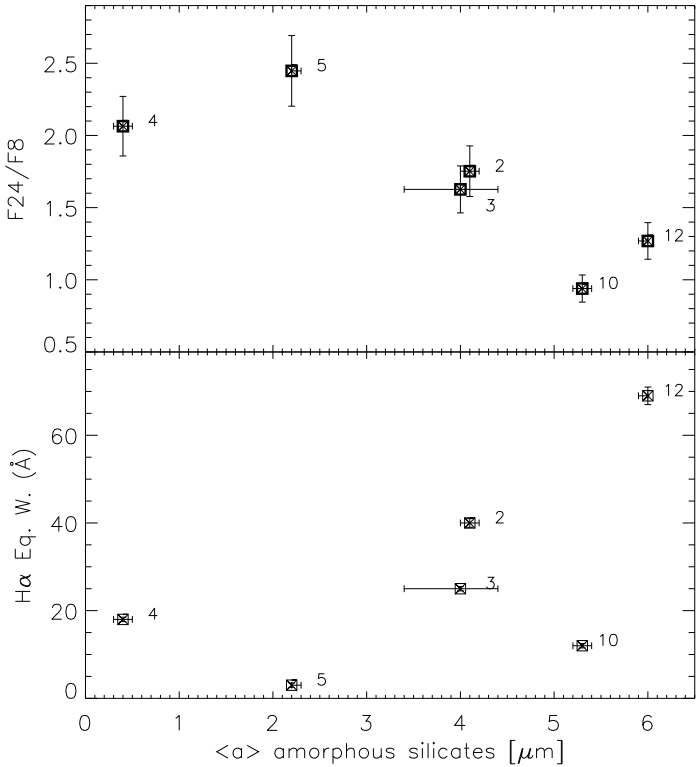}
   \caption{Disc properties related to the amorphous grain size, derived from
     the warmer region. Top panel: Degree of flaring (derived from the flux
   ratio at 24 and 8~\mic) versus the mass-averaged grain size. The larger the
   grains, the smaller the flaring, what points to dust settling when grains
   have grown. Lower panel: Relation between the equivalent width of H$\alpha$
   and the mass-averaged size of the amorphous grains. The trend suggests
   that, the larger the accretion rate, the larger the grains in the disc
   atmosphere are.}
         \label{flaringsize}
   \end{figure}

The classical model of a T Tauri disc is one in which the optically thick
midplane (containing the bulk of the disc mass) is surrounded by an optically
thin disc atmosphere, which is hot and source of the infrared emission
features. The disc can flare under hydrostatical equilibrium, and when the gas
and dust are well-mixed, the turbulent motion of the gas will support the dust
grains in the disc atmosphere against gravitational settling towards the disc
midplane. In the disc evolution model from Dullemond \& Dominik
(\cite{dullemond2004}), grains located in the atmosphere of a flared disc that
grow can no longer be supported by the gas. As a consequence, the larger
grains will settle towards the disc midplane, hereby reducing the amount of
flaring in the disc. In Fig.~\ref{flaringsize}, we plot the ratio of the
fluxes at 24 and 8~\mic \ versus the derived grain size of the amorphous
silicates. This flux ratio can be used as a proxy for the degree of
flaring in a disc (e.g. Apai et al.\ \cite{apai2005}, Scholz et al.\
\cite{scholz2007} for brown dwarfs and Furlan et al.\ \cite{furlan2005} for
TTS): the larger the ratio, the more flared the disc. It is clear that the
more flared the MBM 12 discs are, the smaller the derived silicate grain
sizes, in agreement with the dust settling model from Dullemond \& Dominik
(\cite{dullemond2004}).

\subsubsection{Accretion rate}


We only have two weak-line T Tauri stars in our {\it detected} sample: MBM
12-1, which has no excess emission, and MBM 12-10, which has an IR excess due
to warm dust. We thus see accretion in all but one of the objects which have
an IR excess. Sicilia-Aguilar et al.\ (\cite{aurora2007}) showed a particular
relation between the accretion rate derived from the U-band emission, and
the grain sized derived from the 10~\mic \ feature: they found that objects
with higher accretion rates have, in general, larger dust grains. This suggest
that turbulence supports large dust grains against settling, and that when
the accretion weakens, it will become apparent in the observed grain size. In
Fig.~\ref{flaringsize}, we plot the equivalent width of the \halpha \ line in
function of the derived grain size, and we find a similar tendency for our T
Tauri stars - apart from object MBM 12-10, which is a WTTS. It is thus
tempting to assume that, the larger turbulence is, the easier it is to support
large grains against settling towards the disc midplane. Caution should be
taken, however, given 1) the variability of the \halpha \ line, and 2) the
spectral dependency of the equivalent width for a given accretion
rate. Furthermore, some weak-line TTS - based on the \halpha \ equivalenth
width - were found to anyway be accreting, through the analysis of the width
of their broad \halpha \ line (Sicilia-Aguilar et al.\ \cite{aurora2006}).
Therefore, additional studies, where the line is observed at several
epochs and with higher spectral resolution (so that the accretion rate can be
derived from the 10~\% velocity width instead of the equivalent width) are 
required to better understand this relation.


\section{Conclusions}
\label{conc}

We presented {\it Spitzer} spectroscopy and photometry for the complete sample
of T Tauri stars in the 2 Myr-old star forming cloud MBM 12. From the 12
objects observed, 8 were detected and could be studied more in detail in this
paper.  We composed spectral energy distributions for our sample, and found
that only 1 object shows no excess at all, and is probably a disc-less T Tauri
star. The other objects show excesses that can be attributed to the presence
of a protoplanetary disc. Of the 7 IR excess sources, 3 have candidate
transitional discs.

We analysed the properties of the warm dust (visible through their solid-state
bands in the infrared) with the two-layer temperature distribution method from
Juh\'asz et al.\ (\cite{juhasz2008}) and derived the mass fraction of the
different species, and their mass-averaged sizes. We showed that the dust
components seen in other young stars can very well explain the spectral
features observed in MBM 12 discs. We also found evidence for grain growth:
the shape and the strength of the 10~\mic \ feature shows that grains with 
submicron sizes, as well as grains of sizes up to a few microns are present 
in the discs in varying amounts. The presence
of crystalline silicates demonstrate that high-temperature processes take
place early around other stars, consistent with meteoritic evidence from the
young solar system. In a next step, we related the derived properties with the
stellar and disc properties, and found:

\begin{enumerate}
\item The later the spectral type of the central object, the weaker the
  10~\mic \ feature, implying larger dust grains. This is most likely a
  luminosity effect, as in less luminous stars, the radial distance observed
  at 10~\mic \ is smaller, hence the density larger and, as a consequence, 
  faster grain growth - it does not need to imply a real difference in the 
  size distribution
\item A spatial gradient in grain growth: the mass-averaged grain sizes are 
      larger in the warmer region than in the cooler region, which also can be
      related to a difference in density
\item There is a large spread in the degree of crystallisation (between 3 and $>$
  30\%), despite a similar age for the objects; the crystallisation fraction
  is independent of spectral type
\item The crystallisation is also independent of grain growth: the mass 
  fraction in crystalline grains nor the size of the crystalline grains are
  correlated with the size of the amorphous silicates. However, the largest
  crystalline fractions are observed for those sources with largest grain sizes
\item A spatial gradient in crystalline composition, as observed in
  the forsterite to enstatite mass ratio: forsterite is much more dominating
  in the outer than in the inner disc region. Enstatite is more abundant in
  those sources with larger grains, and in regions where the density is higher. This
  suggests that its formation process needs higher densities than forsterite 
\item The presence of a spatial gradient in both grain sizes and composition
  indicates that radial mixing is not a very efficient process in these
  protoplanetary discs
\item Companions, even at distances larger than 70 AU, influence disc
  evolution: the IR excess of single sources is a factor 3 higher than that of
  binaries. 
\item Disc flaring and grain size are mildly related: the more the discs
  flare, the smaller the grains in their atmosphere
\item There is tendency for objects with a stronger \halpha \ line to have
  larger grains, suggesting that more heavily accreting (more turbulent)
  sources can support larger grains in their disc atmosphere more
  efficiently 

\end{enumerate}

We are aware that these findings are based on a small sample, and that it
should be confronted with a larger database of well-known T Tauri stars, for
which infrared spectra of high quality are available. This study is part of a
larger programme with {\it Spitzer} on discs in young clusters. The low-mass
M-type stars in the Coronet cluster have already been discussed in
Sicilia-Aguilar et al.\ (\cite{aurora2008}), while T Tauri stars in the $\eta$
Cha cluster will be presented by Sicilia-Aguilar et al.\ ({\it in
preparation}) and T Tauri stars in the $\epsilon$ Cha cluster by Fang et al.\
({\it in preparation}). From a comparison of the results in the different
clusters, we will be able to better determine the relations between the
different stellar and dust properties, so that they can be incorporated in
theoretical (evolutionary) disc models of young objects.


\begin{acknowledgements}

It is a pleasure to thank Ray Jayawardhana, who sparked GM's interest in this
cloud, Leen Decin for calculating the appropriate MARCS models for our sample 
and Jeff Nichols for help with Python. We also thank the referee, Dan Watson,
for his prompt report. GM and AS-A acknowledge support by
the \emph{Deut\-sche For\-schungs\-ge\-mein\-schaft, DFG\/}, project numbers
ME 2061/3-2 and SI 1486/1-1, respectively. WL asknowledges support from the 
Australian  Academy of Sciences international exchange program. We made 
extensively use of the SAO/NASA Astrophysics Data System and SIMBAD hosted by 
CDS, Strasbourg.

\end{acknowledgements}


\section*{Appendix}

\begin{table*}
\begin{center}
\caption{Parameters involved in the TLTD model}
\label{tparam}
\begin{tabular}{ll}       
\hline\hline\noalign{\smallskip}
Parameter & Meaning\\
\noalign{\smallskip}
\hline\noalign{\smallskip}
$F_\nu$ & Observed flux \\
$F_{\nu, cont}$  & Total continuum flux  \\
$d$ & Distance to the source\\
$R_*$ & Stellar radius\\
$T_*$ & Stellar temperature \\
$D_0$ & Contribution of the star to the total flux\\
$D_1$ & Contribution of the inner rim to the total flux \\
$D_2$ & Contribution of the midplane to the total flux \\
$D_{i,j}$       & Mass contribution of species i with size j \\
$T_{a,\ \rm out}$ & Lowest temperature in the disk atmosphere \\
$T_{a,\ \rm in}$ & Highest temperature in the disk atmosphere \\
$T_{r,\ \rm out}$ & Lowest temperature in the inner rim \\
$T_{r,\ \rm in}$ & Highest temperature in the inner rim \\
$T_{m,\ \rm out}$ & Lowest temperature in the disk midplane\\
$T_{m,\ \rm in}$ & Highest temperature in the disk midplane \\
$\kappa _{i,j}$ & Mass absorption coefficient of species i with size j \\
$qr$ & Power exponent of the temperature distribution (as a function of radius) 
in the rim \\
$qa$ & Power exponent of the temperature distribution (as a function of radius) 
in the disk atmosphere \\
$qm$ & Power exponent of the temperature distribution (as a function of radius) 
in the disk midplane \\
\noalign{\smallskip}
\hline
\end{tabular}
\end{center}
\end{table*}

\begin{table*}
\caption{For each object, we list the reduced $\chi^{2}$ of the fit, and the
  mass fraction of each species for the 3 different sizes, used to fit the
  shorter and longer wavelengths regions (upper and lower part of the table,
  respectively). We only show mass fractions that are higher than 5\%.}
\label{tfit}     
\centering                        
\begin{tabular}{lcccccccc}       
\hline\hline\noalign{\smallskip}
        &                &MBM 12-2   &MBM 12-3    &MBM 12-4   &MBM 12-5&MBM 12-6&MBM 12-10&MBM 12-12 \\
\noalign{\smallskip}
\noalign{\smallskip}
\hline\noalign{\smallskip}
7-17 micron range & $\chi^{2}$ &12.2   &5.7     &9.0   &9.7         &4.9        &3.4      &5.6  \\
\noalign{\smallskip}
\hline\noalign{\smallskip}
Species &Size&\multicolumn{7}{c}{Mass fractions (in percentage)}\\
\noalign{\smallskip}
\hline\noalign{\smallskip}
Amorphous silicate   & 0.1~\mic   & 13.2 ($^{+0.4}_{-0.4}$)   & --                        & 48.2 ($^{+2.3}_{-2.1}$)   & 43.4 ($^{+0.4}_{-0.4}$)   & --                        & --                        & --                        \\
(Olivine Type)       & 1.5~\mic   & --                        & --                        & --                        & --                        & --                        & --                        & --                        \\
                     & 6.0~\mic   & --                        & 29.7 ($^{+13.1}_{-14.2}$) & --                        & --                        & 8.8 ($^{+27.6}_{-8.7}$)   & 17.6 ($^{+29.2}_{-15.7}$) & 53.3 ($^{+7.5}_{-9.5}$)   \\
Amorphous silicate   & 0.1~\mic   & --                        & --                        & 15.9 ($^{+3.9}_{-2.8}$)   & --                        & --                        & --                        & --                        \\
(Pyroxene Type)      & 1.5~\mic   & 20.2 ($^{+0.8}_{-0.8}$)   & 35.2 ($^{+9.7}_{-7.9}$)   & 16.8 ($^{+3.2}_{-3.3}$)   & 24.0 ($^{+0.6}_{-0.9}$)   & --                        & 13.8 ($^{+3.3}_{-3.3}$)   & --                        \\
                     & 6.0~\mic   & 55.7 ($^{+0.9}_{-1.0}$)   & 19.8 ($^{+18.6}_{-12.9}$) & --                        & 29.0 ($^{+1.4}_{-1.2}$)   & 32.0 ($^{+34.3}_{-30.5}$) & 62.0 ($^{+16.0}_{-28.4}$) & 10.9 ($^{+9.7}_{-8.3}$)   \\
Forsterite           & 0.1~\mic   & --                        & --                        & --                        & --                        & --                        & --                        & --                        \\
                     & 1.5~\mic   & --                        & --                        & --                        & --                        & --                        & --                        & --                        \\
                     & 6.0~\mic   & --                        & --                        & --                        & --                        & --                        & --                        & --                        \\
Enstatite            & 0.1~\mic   & --                        & --                        & --                        & --                        & --                        & --                        & --                        \\
                     & 1.5~\mic   & --                        & --                        & --                        & --                        & --                        & --                        & --                        \\
                     & 6.0~\mic   & 9.0 ($^{+0.8}_{-0.9}$)    & --                        & 10.6 ($^{+3.1}_{-3.2}$)   & --                        & 40.1 ($^{+19.7}_{-13.1}$) & --                        & 29.1 ($^{+2.6}_{-2.6}$)   \\
Silica               & 0.1~\mic   & --                        & --                        & --                        & --                        & --                        & --                        & --                        \\
                     & 1.5~\mic   & --                        & --                        & --                        & --                        & --                        & --                        & --                        \\
                     & 6.0~\mic   & --                        & 7.0 ($^{+2.9}_{-2.2}$)    & --                        & --                        & 5.4 ($^{+6.6}_{-4.6}$)    & --                        & --                        \\
\noalign{\smallskip}
\hline\noalign{\smallskip}
17-37 micron range& $\chi^{2}$ &12.3    &8.3     &14.0    &14.2    &5.9     &3.6      &5.7\\
\noalign{\smallskip}
\hline\noalign{\smallskip}
Species &Size&\multicolumn{7}{c}{Mass fractions (in percentage)}\\
\noalign{\smallskip}
\hline\noalign{\smallskip}
Amorphous silicate   & 0.1~\mic   & 65.2 ($^{+2.9}_{-4.7}$)   & --                        & 51.1 ($^{+5.6}_{-9.6}$)   & --                        & --                        & 37.6 ($^{+13.1}_{-15.4}$) & 30.5 ($^{+17.0}_{-22.5}$) \\
(Olivine Type)       & 1.5~\mic   & --                        & 68.8 ($^{+3.8}_{-4.6}$)   & 6.4 ($^{+10.4}_{-5.3}$)   & 36.7 ($^{+2.6}_{-4.9}$)   & --                        & --                        & 32.7 ($^{+19.0}_{-14.3}$) \\
                     & 6.0~\mic   & --                        & --                        & --                        & --                        & --                        & --                        & --                        \\
Amorphous silicate   & 0.1~\mic   & 30.4 ($^{+3.6}_{-4.2}$)   & 21.0 ($^{+3.6}_{-6.2}$)   & 33.0 ($^{+1.6}_{-3.2}$)   & 47.6 ($^{+3.8}_{-13.6}$)  & 8.1 ($^{+58.2}_{-7.9}$)   & 41.7 ($^{+12.8}_{-14.4}$) & 31.5 ($^{+4.5}_{-6.2}$)   \\
(Pyroxene Type)      & 1.5~\mic   & --                        & --                        & --                        & --                        & 79.4 ($^{+10.0}_{-31.6}$) & --                        & --                        \\
                     & 6.0~\mic   & --                        & --                        & --                        & --                        & --                        & --                        & --                        \\
Forsterite           & 0.1~\mic   & --                        & --                        & --                        & --                        & 5.8 ($^{+0.9}_{-0.7}$)    & --                        & --                        \\
                     & 1.5~\mic   & --                        & --                        & --                        & --                        & --                        & --                        & --                        \\
                     & 6.0~\mic   & --                        & --                        & --                        & --                        & --                        & --                        & --                        \\
Enstatite            & 0.1~\mic   & --                        & --                        & --                        & --                        & --                        & --                        & --                        \\
                     & 1.5~\mic   & --                        & --                        & --                        & --                        & --                        & --                        & --                        \\
                     & 6.0~\mic   & --                        & --                        & --                        & --                        & --                        & --                        & --                        \\
Silica               & 0.1~\mic   & --                        & --                        & --                        & --                        & --                        & --                        & --                        \\
                     & 1.5~\mic   & --                        & --                        & --                        & --                        & --                        & --                        & --                        \\
                     & 6.0~\mic   & --                        & --                        & 6.4 ($^{+0.9}_{-1.1}$)    & 7.1 ($^{+0.6}_{-0.5}$)    & --                        & 8.7 ($^{+4.7}_{-4.3}$)    & --                        \\
\noalign{\smallskip}
\hline
\end{tabular}       
\end{table*}

\end{document}